\documentclass[11pt,a4paper,twoside]{article}

\usepackage{amssymb}
\usepackage{amsmath}
\usepackage{bm}
\usepackage{graphicx}
\usepackage{subfigure}
\usepackage{hyperref}
\usepackage{color}
\usepackage{revsymb}
\usepackage[font=small,labelfont=bf]{caption}
\usepackage[title]{appendix}

\usepackage{newtxmath}
\usepackage{booktabs}
\usepackage{array}
\usepackage{enumitem}
\usepackage{cite}
\usepackage[affil-it,auth-lg]{authblk}
\usepackage{hyperref}
\usepackage{eso-pic}

\usepackage{lineno}

\newcommand{\e}{{({\rm e})}}

\newcolumntype{K}[1]{>{\centering\arraybackslash}p{#1}}

\begin{document}

	\title{Thermodynamic instability of dark energy equation of state}
	
	\author[1,2]{Zacharias Roupas}
	\affil[1]{Faculty of Energy and Environmental Engineering, The British University in Egypt, Sherouk City 11837, Cairo, Egypt} 
	\affil[2]{Department of Physics, University of Crete, 70013, Herakleion, Greece} 
	
	\date{\vspace{-5ex}}
	
	
	\maketitle

\begin{abstract}
	We {derive} the dark energy fluid equation of state $P = -\epsilon = {\rm const.}$ as an extremum of entropy, subject to the Hamiltonian constraint of General Relativity. However, we identify perturbations that can render this extremum an entropy minimum designating a thermodynamic instability and specify the mathematical condition for this to occur. 
	
\end{abstract}

\section{Introduction}

The accelerated expansion of the universe \cite{Riess_1998,1999ApJ...517..565P} may be attributed to dark energy, whose nature remains as of yet elusive \cite{Faraoni_2011}. The cosmological constant, being a primary candidate of dark energy, may be of purely geometrical origin or a manifestation of vacuum energy \cite{2003RvMP...75..559P}. Other possibilities are inflationary fields, 
quintessence \cite{1998PhRvL..80.1582C,PhysRevD.37.3406,1999PhRvL..82..896Z}
or phantom \cite{2002PhLB..545...23C}, and modified gravity theories (e.g. see \cite{2016ARNPS..66...95J} and references therein). 

We will work here in the hydrodynamic limit of a dark energy fluid in an attempt to probe the equation of state using entropic arguments, remaining agnostic on the origin and nature of that fluid. {The fluid approximation for perturbations shall be reliable at least in long wavelengths  \cite{1997PhRvD..55.5205F,2000CQGra..17.2983F}. 
We shall not pre-assume an equation of state, but will derive the dark energy equation of state $P = -\epsilon={\rm const.}$ as the entropy extremum of a fluid in an isotropic, spherically symmetric spacetime. 
However, our second order variation reveals perturbations that can render this extremum an entropy minimum, designating a thermodynamic instability. 
}
	
In the next section we {derive the dark energy equation of state from entropy. In section \ref{sec:2ndOP} we perform the second order variation analysis} and conclude our results in the last section.

\section{{Derivation of the dark energy e.o.s. from entropy}}

We shall calculate the entropy extrema on a static spacetime, that is generally expressed with the line element
\begin{equation}\label{eq:metric}
	d\ell^2  = -g_{tt}(r)c^2 dt^2 + g_{rr}(r) dr^2 + r^2d\Omega,
\end{equation}
for, as yet, unspecified functions $g_{tt}$, $g_{rr}$, {assuming that the Hamiltonian constraint of General Relativity holds. We shall assume the presence of a fluid with unknown equation of state.}
  
	For full generality, we remain agnostic on the nature of this fluid, assuming any number of effective fluid particles' species with arbitrary chemical potential. 
We shall use the subscript $q$ to count species. We shall denote the number of particles of the $q$-th specie as $N_q(r)$, its chemical potential as $\mu_q(r)$ and the particles' number density as $n_q(r) = dN_q(r)/dV$. We, therefore, assume that an arbitrary number of species generates the equation of state 
\begin{equation} 
	P = P (\epsilon,\{ n_q \} ) ,
\end{equation}	 
where $P$ is the local pressure and $\epsilon$ the local, total mass--energy density.
{
We wish to identify a thermal equilibrium $P_{({\rm e})}(\epsilon_{({\rm e})},n_{({\rm e})})$, $g_{rr,{({\rm e})}}$, $g_{tt,{({\rm e})}}$ of this system.}

The first law of thermodynamics for entropy spatial density $s(r)$ is expressed locally as 
\begin{equation}\label{eq:1st_law}
	T(r)ds = d\epsilon - \sum_q\mu_q (r) dn_q,
\end{equation}
where we use units $k=1$, so that temperature $T$ is measured in energy units and entropy is dimensionless. 
This equation implies that in the most general case
\begin{equation}\label{eq:s_dep}
	s = s(\epsilon,\{n_q\}).
\end{equation}
{Therefore, $\epsilon$, $\{n_q\}$ are the quantities that shall be perturbed independently. We shall consider linear perturbations $\delta \epsilon$, $\delta n_q$, that shall induce perturbations at any order in $s(\epsilon,\{n_q\})$, $\mu_q(\epsilon,\{n_q\})$, $P(\epsilon,\{n_q\})$, $T(\epsilon,\{n_q\})$.}
The first law (\ref{eq:1st_law}) gives
\begin{equation}\label{eq:1st_law_var}
	T(r) \delta s(r) = \delta \epsilon (r) - \sum_q\mu_q (r) \delta n_q(r).
\end{equation}

Note that we allow for local temperature $T$ and chemical potential $\mu_q$ to depend on the position in equilibrium. 
{It has been emphasized by Tolman \cite{Tolman} that the temperature, measured by a local observer, is not uniform at thermal equilibrium in General Relativity, because of the gravitation of thermal energy. The latter rearranges itself at equilibrium to support its own gravitational attraction \cite{Roupas_2015CQG}}. There is a second equation involving thermodynamic densities, called Euler equation (or sometimes integrated Gibbs--Duhem relation),
\begin{equation}\label{eq:Euler}
	T(r)s(r) = P(r) + \epsilon (r) - \sum_q\mu_q(r) n_q(r).
\end{equation}
In the followings, we sometimes omit denoting the dependence on $r$ of $T$, $s$, $P$, $\epsilon$, $\mu_q$, $n_q$, $g_{tt}$, $g_{rr}$ to ease our notation.

{We further impose the Hamiltonian constraint of General Relativity. Assuming isotropy and spherical symmetry it is written as
	\begin{equation}\label{eq:hamiltonian_constraint}
		^{(3)}\mathcal{R} = \frac{16\pi G}{c^4} \epsilon
		\Leftrightarrow
		\frac{2}{r^2}\frac{d}{dr}\left(r(1-g_{rr}^{-1})\right)
		= \frac{16\pi G}{c^4} \epsilon
	\end{equation}
	which gives
	\begin{equation}\label{eq:g_rr}
		g_{rr} = \left( 1 - \frac{2 G }{r c^4} \int_0^r 4\pi x^2 \epsilon(x) dx \right)^{-1}.
	\end{equation}
This equation constrains that perturbations in $\epsilon$ induce perturbations in $g_{rr}$ through the quantity
\begin{equation}\label{eq:m}
	m(r)  \equiv \frac{1}{c^2} \int_0^r 4\pi x^2 \epsilon(x) dx \Rightarrow
	\delta m(r)  = \frac{1}{c^2} \int_0^r 4\pi x^2 \delta \epsilon(x) dx 
\end{equation}
in the specific manner (\ref{eq:g_rr}).
}

{We consider perturbations within an arbitrary large sphere of radius $R$, which can be of the size of a galaxy cluster up to a Hubble radius. In the latter case we assume perturbations to vanish on the boundary, so as to avoid singularities.} 
The total number of particles of the $q$th specie is 
\begin{equation}\label{eq:N}
	N_q = \int_0^R \,4\pi r^2 n_q \sqrt{g_{rr}}dr .
\end{equation}
Likewise the total entropy of the system within the same radius is 
\begin{equation}\label{eq:S}
	S = \int_0^R 4\pi r^2 s \sqrt{g_{rr}}dr,
\end{equation}
The total mass--energy of the system including the gravitational potential is \cite{Weinberg_1972gcpa.book}
\begin{equation}\label{eq:Mtot}
	E = \int_0^R 4\pi r^2 \epsilon dr. 
\end{equation}
{ 
This suggests that $m(r)$ as in (\ref{eq:m}) is the total mass,
including the gravitational field energy, contained within radius $r$ and therefore for $M=m(R)$ it is 
\begin{equation}
	E=Mc^2\equiv m(R) c^2.
\end{equation}
}

We consider linear perturbations of $\epsilon$, $n_q$ for constant total mass--energy and number of particles, introducing the Lagrange multipliers
\begin{equation}
	\beta = { \rm const.},\;
	a_q = {\rm const.}.
\end{equation}
We have to first order
{\begin{equation}\label{eq:dS_1}
	\delta^1 	S - \beta \delta^1 E + \sum_q a_q \delta^1 N_q = 0,
\end{equation}
}
where the entropy is assumed to be a functional 
\begin{equation}
	S = S[\epsilon,\{n_q\}]
\end{equation}
through Eq. (\ref{eq:s_dep}). 
Note that the Lagrange multipliers shall be specified by the first order variation (\ref{eq:dS_1}).

{We stress, that to first order this variation is equivalent to the variation of the Massieu function of the Helmholtz free energy $J_F \equiv \beta E - S$ assuming constant number of particles, that is
\begin{equation}
	\delta^1 J_F - \sum_q a_q  \delta^1 N_q = 0,
\end{equation}
and of the variation of the Massieu function of Gibbs free energy $J_G \equiv \beta E - S -\sum a_q N_q$ 
\begin{equation}
	\delta^1 J_G = 0.
\end{equation}
Thus, the thermal extremum is the same under any conditions; open or closed systems. This corresponds to the equivalence of statistical ensembles to first order.
}

{However, to second order the statistical ensembles in gravity are not equivalent \cite{Campa_2014}. This means that the thermal stability properties of self-gravitating systems depend on the boundary conditions. We shall consider second order variations and discuss this issue further in the next section.}

We are now ready to calculate equilibrium suggested by the first order variation in entropy (\ref{eq:dS_1}). Applying the first law of thermodynamics (\ref{eq:1st_law}), we get that{ 
\begin{equation}
	\int_0^R dr\, 4\pi r^2 \left\lbrace 
	\left( \frac{\sqrt{g_{rr}}}{T} - \beta\right)\delta \epsilon +
	\sqrt{g_{rr}} \sum_q\left(a_q - \frac{\mu_q}{T}\right)\delta n_q 
	\left( s + \sum_q a_q n_q\right) g_{rr}^{3/2} \frac{G}{rc^2} \delta m	
	\right\rbrace
	= 0,
\end{equation}}
Since $\delta n_q$ are varied independently we deduce
\begin{equation}\label{eq:mu_T}
	a_q \equiv \frac{\mu_q(r)}{T(r)}, \quad \frac{da_q}{dr} = 0
\end{equation}
for any specie $q$. 
{This is a general relation, specifying the values of the Lagrange multipliers $a_q$. It holds also for the perturbations about the equilibrium equation of state, which we have not as yet specified.
We get substituting (\ref{eq:m}), (\ref{eq:mu_T}) and using Euler Eq. (\ref{eq:Euler})
\begin{equation}
	\int_0^R dr\, 4\pi r^2 
	\left( \frac{\sqrt{g_{rr}}}{T} - \beta\right)\delta \epsilon +
	\int_0^R \left(\int_0^r\, \frac{G}{c^4} (4\pi)^2  \frac{P(r)+\epsilon(r)}{T(r)} g_{rr}(r)^{3/2} r x^2 \delta\epsilon (x) dx \right)dr  	
	= 0.
\end{equation}
We perform a change of sequence of integration. For any functions $f_1$, $f_2$, it is
\begin{equation}
	\int_0^R \left(\int_0^r\, f_1(r) f_2(x) dx \right)dr =
	\int_0^R \left(\int_x^R\, f_1(r) f_2(x) dr \right)dx =  
	\int_0^R \left(\int_r^R\, f_1(x) f_2(r) dx \right)dr,  
\end{equation}
where in the final step we simply renamed the variables $r\leftrightarrow x$. Therefore the first order variation is written finally as 
\begin{equation}
	\int_0^R dr\, 4\pi r^2 \delta\epsilon(r)
	\left\lbrace \frac{\sqrt{g_{rr}(r)}}{T(r)} - \beta +
	\int_r^R \frac{4\pi G}{c^4} \frac{P(x)+\epsilon(x)}{T(x)} g_{rr}(x)^{3/2} x dx   	
	\right\rbrace 
	= 0.
\end{equation}
}

{
This equation allows us to specify also the general value of the Lagrange multiplier $\beta$, that is
\begin{equation}\label{eq:beta}
	\beta \equiv \frac{\sqrt{g_{rr}(r)}}{T(r)} +
\int_r^R \frac{4\pi G}{c^4} \frac{P(x)+\epsilon(x)}{T(x)} g_{rr}(x)^{3/2} x dx  
,
\quad
\frac{d\beta}{dr} = 0.
\end{equation}
The quantity $\beta^{-1}$ is called the Tolman temperature. This is the quantity that is uniform at thermal equilibrium in General Relativity, instead of local temperature $T(r)$.}

{Now, we can identify one (not unique) equilibrium equation of state
\begin{equation}\label{eq:Pe}
	P_\e = -\epsilon_\e
\end{equation}
with 
\begin{equation}\label{Tolman_e}
	\beta_\e = \frac{\sqrt{g_{rr,\e}}}{T_\e}.
\end{equation}
We may calculate $\epsilon_\e$ as well as $g_{rr,\e}$, $g_{tt,\e}$ as follows.
}

{The Euler relation (\ref{eq:Euler}), at the equilibrium equation of state (\ref{eq:Pe}) gives
\begin{equation}\label{eq:Euler_eq}
	s_\e = - \sum_q \frac{\mu_{q,\e}}{T_\e}n_{q,\e}\equiv - \sum_q a_{q,\e} n_{q,\e}.
\end{equation}
We take the derivative with respect to $r$, denoted with a prime,
\begin{equation}
	s_\e^\prime = - \sum_q a_{q,\e} n_{q,\e}^\prime
\end{equation}
and incorporate the first law of thermodynamics in the form 
\begin{equation} 
s_\e^\prime = \beta_\e \epsilon_\e^\prime - \sum_q a_{q,\e} n_{q,\e}^\prime 
\end{equation} 
to get 
\begin{equation} 
	\epsilon_\e^\prime = 0 .
\end{equation} 
The equilibrium equation of state is consequently
\begin{equation}\label{eq:eos_e}
	P_\e = -\epsilon_\e = {\rm const.}.
\end{equation} 
We get directly the spatial component of the metric (\ref{eq:g_rr}) at the equilibrium e.o.s.
\begin{equation}
	g_{rr,\e} = \frac{1}{1 - \frac{2G}{c^4}\int_0^r 4\pi x^2 \epsilon_\e dx} = \frac{1}{1- \frac{8\pi G \epsilon_\e}{3c^4} r^2}.
\end{equation}
Thus, we identify $\epsilon_\e$ as the dark energy density, and the radial component of the metric (\ref{eq:metric}) at the equilibrium e.o.s. (\ref{eq:eos_e}) as the de Sitter radial metric component with a horizon 
\begin{equation}
	r_{\rm H} = \sqrt{\frac{3c^4}{8\pi G \epsilon_\e}} .
\end{equation}
}

{We verify that our entropic framework reproduces also the correct de Sitter time component of the metric as follows.
We get by differentiation of the general definition of $\beta$ that in general (even for perturbed quantities about the equilibrium e.o.s.) holds
\begin{equation}
	\frac{d\ln T}{dr} = -g_{rr} \left(\frac{Gm}{r^2c^2} + \frac{4\pi G}{c^4}P\cdot r \right).
\end{equation}
Integrating we get
\begin{equation}
	T(r) = T(R)\exp{ \left\lbrace \int_r^R dx\, g_{rr} \left(\frac{Gm}{x^2c^2} + \frac{4\pi G}{c^4}P\cdot x \right) \right\rbrace} .
\end{equation}
Comparing with the Tolman relation (note that Tolman originally derived this equation as an entropy extremum \cite{Tolman})
\begin{equation}
	T(r) \sqrt{g_{tt}(r)} =\beta^{-1},
\end{equation}
allows us to identify $g_{tt}$ as
\begin{equation}
	g_{tt}(r) = g_{rr}(R)^{-1} 
	\exp{ \left\lbrace -2\int_r^R dx \, g_{rr} \left(\frac{Gm}{x^2c^2} + \frac{4\pi G}{c^4}P\cdot x \right) \right\rbrace } 
\end{equation}
At the equilibrium e.o.s.  (\ref{eq:eos_e}), this gives
\begin{align*}
	g_{tt,\e}(r) &=  \left(1 - \frac{8\pi G}{3c^4}R^2 \right) 
	\exp{ \left( -\int_r^R dx \, 
		\frac{-\frac{8\pi G \epsilon_\e}{3c^4} 2x}{1-\frac{8\pi G \epsilon_\e}{3c^4} x^2}
		 \right)} 
	 =
	 \left(1 - \frac{8\pi G}{3c^4}R^2 \right) 
	 \exp{ \left( -\ln \frac{ 1-\frac{8\pi G \epsilon_\e}{3c^4} R^2 }{ 1-\frac{8\pi G \epsilon_\e}{3c^4} r^2 }
	 	\right)} 
 	\\
 	&= 1-\frac{8\pi G \epsilon_\e}{3c^4} r^2 = g_{rr,\e}(r)^{-1}.
\end{align*}
Thus, we were able to derive the de Sitter space 
\begin{equation}
	d\ell_\e^2 = - g_{tt,\e} c^2 dt^2 + g_{rr,\e} dr^2 + r^2 d\Omega,
\end{equation}
along with the equation of state $P_\e=-\epsilon_\e={\rm const}$, as the thermal equilibrium state of a static, spherically symmetric spacetime, subject to the Hamiltonian constraint of General Relativity. 
}

\section{Second order variation}\label{sec:2ndOP}

We shall perform now a second order variation of entropy {within the context of General Relativity and identify unstable perturbations. Let us remark that proving stability of a system is a much harder task than proving instability. The reason is that you need to account for all possible perturbations to prove stability, while you need just one physically viable unstable mode to prove the instability of a system without loss of generality. Here, we shall identify a class of perturbations that render the dark energy fluid equation of state (\ref{eq:eos_e}) unstable.}

{
The linear perturbations about the equilibrium e.o.s. in the independent variables $\epsilon$, $n_q$
\begin{equation}
	\epsilon = \epsilon_\e + \delta\epsilon,\quad
	n_q = n_{q,\e} + \delta n_q 
\end{equation}
induce second order variations of the entropy $S$ and the number $N_q$ through the second order variations of entropy density $s$ and the metric function $\sqrt{g_{rr}}$. In particular, it is
\begin{align}
\label{eq:d2_s}	\delta^2 s &= -\frac{1}{T_\e}\delta T \delta \epsilon - \sum_q \delta a_q \delta n_q, \\
\label{eq:d2_g}	\delta^2 \sqrt{g_{rr}} &= \frac{3G}{rc^2}g_{rr,\e}^{5/2}(\delta m)^2,
\end{align}
where
\begin{equation}
	T = T_\e + \delta T,\quad
	a_q = a_{q,\e} + \delta a_q .
\end{equation}
The perturbation in the local temperature is related to the perturbation of the Lagrange multiplier
\begin{equation}
	\beta = \beta_\e+ \delta\beta 
\end{equation}
by the general relation (\ref{eq:beta}), which gives
\begin{equation}\label{eq:delta_beta}
	\delta\beta = - g_{rr,\e}^{1/2}\frac{1}{T_\e}\delta T +
	\frac{1}{T_\e} g_{rr,\e}\frac{G}{rc^2}\delta m + 
	\int_r^R\beta_\e \frac{4\pi G}{c^4}(\delta P(x) + \delta \epsilon(x)) g_{rr,\e}xdx.
\end{equation}
The perturbation relation $\delta P=\delta P(\delta \epsilon,\{\delta n_q\})$ with
\begin{equation}
	P = P_\e+\delta P
\end{equation}
is unspecified, but is subject to relativistic and thermodynamic constraints. In particular, we have from the first law of thermodynamics and Euler relation that in general
\begin{equation}
	\left. 
	\begin{array}{l}	
\displaystyle		s^\prime = \frac{1}{T}\epsilon^\prime  - \sum_q a_q n_q^\prime 
		\\
\displaystyle		s = \frac{P+\epsilon}{T} - \sum_q a_q n_q		
	\end{array} 
\right\rbrace 
P^\prime = (P+\epsilon) (\ln T)^\prime .
\end{equation}
This relation imposes the following constraint on $\delta P$ and $\delta\epsilon$ 
\begin{equation}
	(\delta P)^\prime = (\delta P + \delta \epsilon) (\ln T_\e)^\prime
\end{equation}
which allows us to calculate the integral in (\ref{eq:delta_beta}).
Differentiating $\beta_\e=\frac{\sqrt{g_{rr,\e}}}{T_\e}$ we get
\begin{equation}
	(\ln T_\e)^\prime = \frac{8\pi G \epsilon_\e}{3c^4} g_{rr,\e}\cdot r.
\end{equation}
and therefore
\begin{equation}
\frac{4\pi G\beta_\e}{c^4} (\delta P + \delta \epsilon)	g_{rr,\e}\cdot r = \frac{3\beta_\e}{2\epsilon_\e}(\delta P)^\prime .
\end{equation}
Substituting into (\ref{eq:delta_beta}) we get
\begin{equation}
	\delta \beta = -\beta_\e \frac{\delta T}{T_\e} + \beta_\e \frac{G}{rc^2} g_{rr,\e}\delta m +
\frac{3\beta_\e}{2\epsilon_\e}\left( \delta P(R) - \delta P(r) \right).
\end{equation}
}

{
The perturbation $\delta P$ is also constrained by perturbations $\delta a_q$. We have that $s = \frac{P+\epsilon}{T} - \sum_q a_q n_q$ gives	
\begin{equation}
	\delta s = -\frac{\delta P + \delta \epsilon}{T_\e} 
	-\sum_q n_{q,\e} \delta a_q - \sum_q a_{q,\e} \delta n_q .
\end{equation}
By substitution of the first law 
\begin{equation}\label{eq:1st_law_var-e}
\delta s = \frac{1}{T_\e}\delta \epsilon - \sum_q a_{q,\e} \delta n_q
\end{equation} 
we get the constraint
\begin{equation}\label{eq:deltaP_con}
	\delta P = -T_\e \sum_q n_{q,\e}\delta a_{q,\e} . 
\end{equation}
}

The second order variation of $N_q$ is
\begin{equation}
	\delta^2 N_q = \int_0^{R} \left(\delta n_q \delta(\sqrt{g_{rr}}) + n_q \delta^2 (\sqrt{g_{rr}}) \right) 4\pi r^2 dr.
\end{equation}
In both canonical and microcanonical ensembles, $N_q$ is conserved $\delta^2N_q = 0$, which gives a constraint on the types of variations
\begin{equation}\label{eq:var_constraint}
	\delta n_q \delta(\sqrt{g_{rr}}) + n_q \delta^2 (\sqrt{g_{rr}}) = 0.
\end{equation}

{
	We are able now to calculate the second order variation of entropy. It is
\begin{equation}\label{eq:delta2S_1} 
	\delta^2 S = \int_0^{R} dr\,  4\pi r^2 
	\left( s_\e\delta^2\sqrt{g_{rr}} + \delta s \delta(\sqrt{g_{rr}}) + \sqrt{g_{rr}} \delta^2 s \right)
\end{equation}
Invoking the first law (\ref{eq:1st_law_var-e}), the Euler relation at equilibrium (\ref{eq:Euler_eq}), the equilibrium equation of state (\ref{eq:eos_e}) and the constraint (\ref{eq:var_constraint}) we get 
\begin{equation}\label{eq:delta2S_2} 
	\delta^2 S = \int_0^{R} dr\,  4\pi r^2 
	\left( -\beta_\e\frac{1}T_{\e} \delta T\delta\epsilon + \frac{G}{rc^2}\frac{1}{T_\e}g_{rr}^{3/2} \delta m \delta \epsilon  + \sqrt{g_{rr,\e}} \sum_q \delta a_q \delta n_q\right) .
\end{equation}
}

{Let us consider perturbations that preserve the Tolman temperature, that is equivalent to working in the canonical ensemble. This is physically appropriate for systems such as a galaxy cluster or a Hubble patch of the Universe, that may exchange heat with their environment. From the canonical condition
\begin{equation}\label{eq:db_T}
	\delta \beta = 0
\end{equation}
we get
\begin{equation}\label{eq:delta_T}
	\delta T = \beta_\e^{-1}\left( \frac{G}{rc^2g_{rr,\e}^{3/2}}\delta m + \frac{3}{2\epsilon_\e} g_{rr,\e}^{1/2}\left( \delta P(R) - \delta P(r) \right) \right).
\end{equation}
Note that $\delta T(R)\propto \delta E$. If the system gains mass--energy the boundary temperature is increased.
}

{
	Without loss of generality (since we do not prove stability and since $\delta a_q$ are independent) we consider the simplifying assumption that all species are subject to the same perturbation
	\begin{equation}
		\delta a_q=\delta a.
	\end{equation}
Finally, substituting (\ref{eq:deltaP_con}) and (\ref{eq:delta_T}) into the variation (\ref{eq:delta2S_2}) we get 
\begin{equation}\label{eq:delta2S} 
	\delta^2 S_C = \int_0^{R} dr\,  4\pi r^2 \beta_\e 
	\left(
	\frac{3}{2\epsilon_\e} \delta\epsilon \left( \delta P(r) - \delta P(R)\right) + \sum_q \frac{1}{n_{q,\e}} \delta P\delta n_q 	\right) ,
\end{equation}
where we use the subscript $C$ to denote that we consider canonical perturbations (\ref{eq:db_T}).
Therefore, although (\ref{eq:delta2S}) is an entropy variation, it describes variations in the canonical ensemble. In the case of a closed system, i.e. in the microcanonical ensemble (which however is unphysical for patches of the Universe such as galaxy clusters),  there appears in (\ref{eq:delta2S}) an additional term $\delta \beta \delta \epsilon$, that nevertheless gets eliminated because of the microcanonical constraint
\begin{equation}\label{eq:mc_con}
	\int_0^Rdr\, 4\pi r^2 \delta \epsilon =0.
\end{equation}
Eventually, we get in the microcanonical ensemble the same expression as in the canonical one
\begin{align}\label{eq:delta2S_M} 
	\delta^2 S_M &= \delta\beta\int_0^{R} dr\,  4\pi r^2 \delta\epsilon  + \int_0^{R} dr\,  4\pi r^2 \beta_\e 
	\left(
	\frac{3}{2\epsilon_\e} \delta\epsilon \left( \delta P(r) - \delta P(R)\right) + \sum_q \frac{1}{n_{q,\e}} \delta P\delta n_q 	\right) 
	\\
	&=
	\int_0^{R} dr\,  4\pi r^2 \beta_\e 
	\left(
	\frac{3}{2\epsilon_\e} \delta\epsilon \left( \delta P(r) - \delta P(R)\right) + \sum_q \frac{1}{n_{q,\e}} \delta P\delta n_q 	\right) = \delta^2S_C \equiv \delta^2 S.
\end{align}
However, note that $S_M$ is subject to the additional constraint (\ref{eq:mc_con}). We discuss this further, below.
}

{
	We shall, in any case, get an instability if
\begin{equation}
	\delta^2 S > 0,
\end{equation}
that is an entropy minimum. In both microcanonical and canonical cases, this condition may be summarized with the same expression
\begin{equation}\label{eq:unstable_modes}
	\frac{3\beta_\e }{2\epsilon_\e}  \int_0^{R} dr\,  4\pi r^2  
	\delta P \delta\epsilon 
	+
	\beta_\e \int_0^{R} dr\,  4\pi r^2 \sum_q \frac{1}{n_{q,\e}} \delta P\delta n_q
	>	
	\frac{\beta_\e }{\epsilon_\e} \delta P(R) \delta E ,
\end{equation}
where $\delta E = 0$ in the microcanonical ensemble, the quantities $\delta \epsilon$, $\delta n_q$ are varied independently and we may impose $\delta P(R)=0$. The relation $\delta P (\delta \epsilon, \delta n_q)$ depends on the nature of the specific fluid under consideration. 
}

{
	Assuming an isobaric perturbation $\delta P(R)=0$ an instability sets in in the cases 
	\begin{align}
		\label{eq:inst_pos}	
&		\beta_\e > 0,
		\quad 
		\frac{3}{2}\int_0^R dr\, r^2 \delta P \delta \epsilon + \int_0^R dr\, r^2 \sum_q \frac{\varepsilon_\e}{n_{q,\e}}  \delta P \delta n_q > 0,
		\\
		\label{eq:inst_neg}	
&		\beta_\e < 0,
		\quad 
		\frac{3}{2}\int_0^R dr\, r^2 \delta P \delta \epsilon  +
		\int_0^R dr\, r^2 \sum_q \frac{\varepsilon_\e}{n_{q,\e}}  \delta P \delta n_q < 0 .
	\end{align}
	The dark energy fluid has positive temperature, when it is a quintessence-like fluid with speed of sound $v_s(r) \leq  c$, i.e. $P(r) \geq -\epsilon_\e$. We also know that the dark energy fluid has negative temperature, when it is a phantom-like fluid \cite{2004NuPhB.697..363G} with speed of sound $v_s(r) \geq c$, i.e. $P(r) \leq -\epsilon_\e$. 
Let us assume further a canonical ensemble, i.e. $\delta E\neq 0$. Such isothermal, isobaric perturbations apply well to systems like galaxy clusters and to any non-isolated patch of the Universe. These perturbations generate an instability in both cases of a dark energy fluid, quintessence-like or phantom-like. 
Consider a quintessence-like fluid, with positive temperature and $\delta P(r) \geq 0$ everywhere, so as $v_s(r) \leq c$. 
Assume that the system absorbs heat $\delta E>0$ with $\delta \epsilon(r) \geq 0$ everywhere. Then, it is subject to the instability case (\ref{eq:inst_pos}). Therefore, a quintessence-like dark energy fluid gets destabilized when it absorbs heat. 
Consider now a phantom-like fluid, with negative temperature and $\delta P(r) \leq 0$ everywhere, so as $v_s(r) \geq c$. 
Assume that the system emits heat $\delta E<0$ with $\delta \epsilon(r) \leq 0$ everywhere. Then, it is subject to the instability case (\ref{eq:inst_neg}). Therefore, a phantom-like dark energy fluid gets destabilized when it emits heat. Note that both fluids may also get destabilized in the reverse cases. We just do not need to prove this. As mentioned before, a single unstable mode is sufficient to prove instability of a system.
}

{
Such perturbations $\delta E\neq 0$ are not allowed in the microcanonical ensemble. Physically, such a strong constraint $\delta E = 0$ at all times may apply only to the Universe as a whole. Still, it is not certain that the thermodynamic instability does not set in even in this case. It is not locally that $\delta P \delta \epsilon$ must be positive, but it is its integral. For example, in the case of an ideal fluid, unstable modes of the canonical ensemble  are indeed suppressed in the microcanonical one, allowing for negative specific heat states to be stable under conditions of constant total energy \cite{Campa_2014} (see also section 4 in \cite{Roupas_JPhysA_2020}). However, the microcanonical ensemble is not totally immune to instability and unstable modes appear at the equilibrium point where the specific heat becomes positive again \cite{Roupas_2015CQG}. In our case, an instability shall set-in in the microcanonical ensemble, for any (and not only) perturbation with $\delta n_{q,\e} = 0$ and $\beta_\e \int_0^R dr\, r^2 \delta P(r) \delta\epsilon (r) > 0$, subject to the constraint (\ref{eq:mc_con}).
}

\section{Conclusion}

	In our approach the equation of state of the dark energy fluid is not assumed to be known, but instead it is expected to be derived as a thermodynamic equilibrium. We manage to derive the equation of state $P_\e = -\epsilon_\e={\rm const.}$ {and the de Sitter space as thermal equilibrium states in a static spherically symmetric spacetime, subject to the Hamiltonian constraint of General Relativity}.

	Our second order variation analysis reveals perturbations that can generate instabilities of this equation of state. 
	{
		We conclude that a thermodynamic instability is ignited by perturbations that satisfy equation (\ref{eq:unstable_modes}). It is evident that the constant density e.o.s. is unstable for both quintessence-like ($v_s(r) \leq c$) and phantom-like ($v_s(r) \geq c$) fluids.
	}

{
	Such instability of dark energy does not mean that the fluid is not physical and therefore non-existent. On the contrary, the thermodynamic instability identified here implies only that the dark energy equation of state departs from the constant value $P_\e=-\epsilon_\e={\rm const}$, generating a local variation of the fluid's speed of sound, which in its turn becomes the source of  gravitational instability. The latter generates structure in the Universe.
	Our calculation provides, therefore, support to the idea that the dark energy fluid, if physical, is subject to clustering \cite{2002PhRvD..65l3505P,2002PhRvD..66d7302A,2011JCAP...11..014A,2015MNRAS.452.2930M,2017JCAP...11..048B}. }

\bibliography{Roupas_DE_inst}

\begin{thebibliography}{10}

\bibitem{Riess_1998}
Adam~G. Riess et~al.
\newblock Observational evidence from supernovae for an accelerating universe
  and a cosmological constant.
\newblock {\em The Astronomical Journal}, 116(3):1009--1038, sep 1998.

\bibitem{1999ApJ...517..565P}
S.~{Perlmutter} et~al.
\newblock {Measurements of {\ensuremath{\Omega}} and {\ensuremath{\Lambda}}
  from 42 High-Redshift Supernovae}.
\newblock {\em \apj}, 517(2):565--586, June 1999.

\bibitem{Faraoni_2011}
Valerio Faraoni.
\newblock Dark energy: Theory and observations.
\newblock {\em Classical and Quantum Gravity}, 28(4):049003, jan 2011.

\bibitem{2003RvMP...75..559P}
P.~J. {Peebles} and Bharat {Ratra}.
\newblock {The cosmological constant and dark energy}.
\newblock {\em Reviews of Modern Physics}, 75(2):559--606, April 2003.

\bibitem{1998PhRvL..80.1582C}
R.~R. {Caldwell}, Rahul {Dave}, and Paul~J. {Steinhardt}.
\newblock {Cosmological Imprint of an Energy Component with General Equation of
  State}.
\newblock {\em \prl}, 80(8):1582--1585, February 1998.

\bibitem{PhysRevD.37.3406}
Bharat Ratra and P.~J.~E. Peebles.
\newblock Cosmological consequences of a rolling homogeneous scalar field.
\newblock {\em Phys. Rev. D}, 37:3406--3427, Jun 1988.

\bibitem{1999PhRvL..82..896Z}
Ivaylo {Zlatev}, Limin {Wang}, and Paul~J. {Steinhardt}.
\newblock {Quintessence, Cosmic Coincidence, and the Cosmological Constant}.
\newblock {\em \prl}, 82(5):896--899, February 1999.

\bibitem{2002PhLB..545...23C}
R.~R. {Caldwell}.
\newblock {A phantom menace? Cosmological consequences of a dark energy
  component with super-negative equation of state}.
\newblock {\em Physics Letters B}, 545(1-2):23--29, October 2002.

\bibitem{2016ARNPS..66...95J}
Austin {Joyce}, Lucas {Lombriser}, and Fabian {Schmidt}.
\newblock {Dark Energy Versus Modified Gravity}.
\newblock {\em Annual Review of Nuclear and Particle Science}, 66(1):95--122,
  October 2016.

\bibitem{1997PhRvD..55.5205F}
J{\'u}lio~C{\'e}sar {Fabris} and J{\'e}r{\^o}me {Martin}.
\newblock {Amplification of density perturbations in fluids with negative
  pressure}.
\newblock {\em \prd}, 55(8):5205--5207, April 1997.

\bibitem{2000CQGra..17.2983F}
J{\'u}lio~C. {Fabris}, Sergio V.~B. {Gon{\c{c}}alves}, and Nazira~A.
  {Tomimura}.
\newblock {An analysis of cosmological perturbations in hydrodynamical and
  field representations}.
\newblock {\em Classical and Quantum Gravity}, 17(15):2983--2998, August 2000.

\bibitem{Tolman}
R.C. Tolman.
\newblock {\em Relativity, Thermodynamics and Cosmology}.
\newblock Oxford, 1934.

\bibitem{Roupas_2015CQG}
Zacharias {Roupas}.
\newblock {Relativistic gravothermal instabilities}.
\newblock {\em Classical and Quantum Gravity}, 32(13):135023, July 2015.

\bibitem{Weinberg_1972gcpa.book}
S.~{Weinberg}.
\newblock {\em {Gravitation and Cosmology}}.
\newblock John Wiley and Sons, July 1972.

\bibitem{Campa_2014}
A.~Campa, T.~Dauxois, D.~Fanelli, and S.~Ruffo.
\newblock {\em Physics of Long-Range Interacting Systems}.
\newblock Oxford University Press, Oxford, 2014.

\bibitem{2004NuPhB.697..363G}
Pedro~F. {Gonz{\'a}lez-D{\'\i}az} and Carmen~L. {Sig{\"u}enza}.
\newblock {Phantom thermodynamics}.
\newblock {\em Nuclear Physics B}, 697(1-2):363--386, October 2004.

\bibitem{Roupas_JPhysA_2020}
Zacharias Roupas.
\newblock Statistical mechanics of gravitational systems with regular orbits:
  rigid body model of vector resonant relaxation.
\newblock {\em Journal of Physics A: Mathematical and Theoretical},
  53(4):045002, Jan 2020.

\bibitem{2002PhRvD..65l3505P}
Francesca {Perrotta} and Carlo {Baccigalupi}.
\newblock {On the dark energy clustering properties}.
\newblock {\em \prd}, 65(12):123505, June 2002.

\bibitem{2002PhRvD..66d7302A}
F.~G. {Alvarenga}, J.~C. {Fabris}, S.~V. {Gon{\c{c}}alves}, and G.~{Tadaiesky}.
\newblock {Clustering of dark energy}.
\newblock {\em \prd}, 66(4):047302, August 2002.

\bibitem{2011JCAP...11..014A}
Stefano {Anselmi}, Guillermo {Ballesteros}, and Massimo {Pietroni}.
\newblock {Non-linear dark energy clustering}.
\newblock {\em \jcap}, 2011(11):014, November 2011.

\bibitem{2015MNRAS.452.2930M}
A.~{Mehrabi}, S.~{Basilakos}, and F.~{Pace}.
\newblock {How clustering dark energy affects matter perturbations}.
\newblock {\em \mnras}, 452(3):2930--2939, September 2015.

\bibitem{2017JCAP...11..048B}
Ronaldo~C. {Batista} and Valerio {Marra}.
\newblock {Clustering dark energy and halo abundances}.
\newblock {\em \jcap}, 2017(11):048, November 2017.

\end{thebibliography}

\bibliographystyle{unsrt}

\end{document}